\documentclass[nolinenumbers,onecolumn]{aastex631}
\usepackage{graphicx}

\submitjournal{PASP}
\accepted{April 22, 2022}

\shorttitle{Ionization in SMC-N66}
\shortauthors{Geist et al.}
\graphicspath{{./}{figures/}}
\begin{document}

%\nolinenumbers

\title{Ionization and Star Formation in the Giant HII Region SMC-N66}

\author[0000-0003-4485-8386]{E. Geist}
\affiliation{Juniata College, 1700 Moore Street, Huntingdon, PA 16652}
\altaffiliation{Department of Astronomy, University of Wisconsin-Madison\\ 475 North Charter St., Madison, WI 53706}

\author[0000-0001-8608-0408]{J. S. Gallagher}
%\affiliation{Department of Astronomy, University of Wisconsin-Madison\\ 475 North Charter St., Madison, WI 53706}

%\collaboration{6}{(AAS Journals Data Editors)}

\author{R. Kotulla}
\affiliation{Department of Astronomy, University of Wisconsin-Madison 475 North Charter St., Madison, WI 53706}

\author{L. Oskinova}

\author{W.-R. Hamann}
\affiliation{Institut f\"ur Physik und Astronomie, Universit\"{a}t Potsdam, Am Neuen Palais 10, House 9, 14469 Potsdam}

\author{V. Ramachandran}
\affiliation{Astronomisches Rechen-Institut, Universit\"{a}t Heidelberg, M\"{o}nchhofstr. 12-14, 69120 Heidelberg}

\author{E. Sabbi}

\author{L. Smith}
\affiliation{Space Telescope Science Institute, 3700 San Martin Dr., Baltimore, MD 21211}

\author{A.Kniazev}
\affiliation{South African Astronomical Observatory, PO Box 9, 7935 Observatory, Cape Town, Republic of South Africa}
\altaffiliation{Southern African Large Telescope, SAAO, SALT, PO Box 9, Observatory Rd 7935 Observatory, Republic of South Africa}

\author{A. Nota}
\affiliation{Space Telescope Science Institute, 3700 San Martin Dr., Baltimore, MD 21211}

\author{M. J. Rickard}
\affiliation{Department of Physics and Astronomy, University College London, Gower Street, London WC1E 6BT, UK}



\begin{abstract}
The NGC~346 young stellar system and associated N66 giant HII region in the Small Magellanic Cloud are the nearest example of a massive star forming event in a low metallicity ($Z\approx0.2Z_{\odot}$) galaxy. With an age of $\lesssim$3~Myr this system provides a unique opportunity to study relationships between massive stars and their associated HII region. Using archival data, we derive a total H$\alpha$ luminosity of L(H$\alpha$)=4.1$\times$10$^{38}$~erg~s$^{-1}$ corresponding to an H-photoionization rate of 3$\times$10$^{50}$~s$^{-1}$. A comparison with a predicted stellar ionization rate derived from the more than 50 known O-stars in NGC~346, including massive stars recently classified from HST FUV spectra, indicates an approximate ionization balance. Spectra obtained with SALT suggest the ionization structure of N66 could be consistent with some leakage of ionizing photons. Due to the low metallicity, the far ultraviolet luminosity from NGC~346 is not confined to the interstellar cloud associated with N66. Ionization extends through much of the spatial extent of the N66 cloud complex, and most of the cloud mass is not ionized. The stellar mass estimated from nebular L(H$\alpha$) appears to be lower than masses derived from the census of resolved stars which may indicate a disconnect between the formation of high and low mass stars in this region. We briefly discuss implications of the properties of N66 for studies of star formation and stellar feedback in low metallicity environments.

\end{abstract}



\section{Introduction} \label{sec:intro}

HII regions are classic signposts of recent star formation. They trace the locations where massive stars form and produce the hydrogen recombination radiation that is a primary measure of star formation rates (SFRs). While the basic treatment of HII regions considers only the effects of ionizing radiation from massive stars\footnote{In this paper "massive stars" refers to stars with masses of $\gtrsim$10~M$_{\odot}$ We distinguish sources of photoionization as being associated with O-stars.}, the structures of HII regions also depend on the impact of stellar winds and, eventually, can be dramatically affected by supernovae. Interactions between HII regions and their surroundings also influence the evolution of the host galaxy, e.g.,  by restructuring the interstellar medium and producing pathways for the escape of ionizing stellar Lyman continuum radiation \citep[see][and references therein]{Choi20}.

Observations of the properties of HII regions in a variety of settings provide essential tests of models of these critical astrophysical systems. The young stellar complex NGC~346 and its associated giant HII region SMC-N66 is
the largest star forming site in the Small Magellanic Cloud (SMC).  Due to the proximity of 
the SMC (D=62~kpc), N66 offers an excellent opportunity to measure the properties of a very 
young HII region in a low metallicity dwarf galaxy. N66 also stands out in terms 
of its large population of O-stars and, as a result of its very young age, absence of supernovae \citep{Reid06, Gouliermis08}. Due to the low metal abundances, stellar winds are weak in main sequence O-stars in the SMC \citep[e.g.,][]{Ramachandran19}. 
Therefore, N66 experienced minimal mechanical feedback from its massive stars \citep{Smith12}, so its large-scale structure is primarily determined by the impact of photoionization on the ambient interstellar matter. The weak winds from main sequence O-stars also reduce the mechanical energy inputs to young H~II regions, thereby affecting their structures.

The opportunities presented by N66 are widely appreciated in the literature.  For example, \cite{Valerdi19} analyzed high quality nebular spectra to derive chemical abundances and estimate the primordial $^4$He abundance. Numerous spectroscopic studies provide a nearly complete census of the OB stellar content \citep[e.g.,][]{Massey89, Evans06, Dufton19}, which have weak stellar winds \citep{Walborn00}. Recently, far-ultraviolet (FUV) Hubble Space Telescope spectra of massive stars in the crowded main NGC~346 star cluster (GO-15112, P.I. L. Oskinova) revealed an additional 7 O-stars (M. Rickard {\it et al.}, in preparation). The ionization rate from OB-stars in NGC~346 is well determined both in terms of stellar identifications and quality of the spectroscopically derived stellar physical parameters. In addition, the bright O-stars enable FUV absorption line studies of the interstellar medium in the SMC, revealing complex velocity and ionization structures \citep[][and references therein]{Danforth02}.

N66/NGC~346 offers an excellent environment to investigate the balance between stellar H-photoionization, nebular recombination rates, and mid-infrared emission \cite[e.g., ][]{Gouliermis06, Gouliermis08}. Nebular ionization rates can be derived from the H$\alpha$ luminosity, L(H$\alpha$).  \cite{Kennicutt86} measured the flux f(H$\alpha$) from photelectrically calibrated narrow-band photographic images. \cite{Caplan96} observed N66 with fixed-aperture photoelectric photometry using a Fabry-Perot to isolate the H$\alpha$ line. Both studies made empirical corrections for the contributions of stars to their measured narrow band fluxes. \cite{Caplan96} also determined the H$\beta$ flux and derived an integrated value for the level of interstellar absorption of 0.3~mag at H$\alpha$, which agrees with spectroscopic measurements \citep{Valerdi19}. A goal of this study is to measure the H$\alpha$ flux of N66 using modern data. 

The issue of Lyman continuum photons escaping from N66 is also of interest. The fraction of escaping ionizing photons is potentially significant for the SMC as a whole since NGC~346 contains a large fraction of the O-star population in this galaxy. \cite{Relano02} compared models for the physical conditions in the nebula with stellar ionization rates to test for escaping ionizing photons. They found a potentially high escape rate of 45\%. On the other hand, \cite{Pellegrini12} compared line ratios of the low ionization potential [SII] lines with [OIII] to search for transition zones where ionization drops in the boundaries of ionization bounded nebulae. Using this approach, they found that N66 is an optically thick nebula with a low escape fraction of ionizing radiation. We revisit this issue using our measurement of L(H$\alpha$), in combination with the more complete census of O-stars in N66 and recent spectroscopic measurements.

Observations of NGC 346 with the HST reveal large populations of stars in pre-main sequence evolutionary phases \citep{Nota06, Simon07, DeMarchi11}. These data have been analyzed to estimate SFRs \citep{Sabbi07, Simon07, Sabbi08, Hony15} that can be compared with those derived from the 
H$\alpha$ and 24~$\mu m$ luminosities \citep{Gouliermis10,Gouliermis14, Hony15}. The stellar mass and star formation history of NGC 346 and its subclumps have been derived from resolved stars and suggest a star formation time scale of $\sim$7~Myr for the main region aside from the older SC16 region \citep{Sabbi07, Cignoni11}, and a total young stellar mass of 2.2$\times$10$^4$~M$_{\odot}$ \citep{Sabbi08, Hony15}. 

Locations of well-defined ionization fronts in N66 have been noted especially through studies of polycyclic aromatic hydrocarbon (PAH) features that were undertaken with the Spitzer Space Telescope (SST) \cite[e.g.,][]{Sandstrom12}.  Dense ionization fronts appear to cover only a fraction of the nebula, which suggests that photoionizing Lyman continuum radiation escapes in some directions into surrounding regions. Unfortunately, as discussed by
\citet{Oey17}, the low dust optical depths in the SMC limit the utility of 8~$\mu m$ PAH emission as an indicator of Lyman continuum optical depths of its HII regions. However, weakly ionization bounded HII regions are potentially important factors in producing the relatively high amounts of diffuse gas in and around the SMC \citep{Smart19} that is a general feature of metal-poor dwarf irregular galaxies \citep[e.g.,][]{Hunter90}.

In this paper, archival narrow-band and mid-infrared images of N66 are used to study the structure of the N66 region and obtain new measurements of H$\alpha$ and mid-infrared fluxes and luminosities. We compare these results with the stellar Lyman continuum luminosity based on a newly-expanded inventory of spectroscopically characterized O-stars.  The possibility for ionizing radiation to escape from N66 is explored based on HII emission line ratios from  \cite{Valerdi19} and spectra we obtained with the Southern African Large Telescope\footnote{Based in part on observations made under proposal 2021-1-SCI-023 with the Southern African Large Telescope (SALT)} (SALT).

\vspace{5mm}
\section{Observations and Luminosities}\label{sec:obs}

\subsection{H$\alpha$ Narrow-Band Photometry}\label{subsec:haptm}

\begin{figure}[ht!]
\plotone{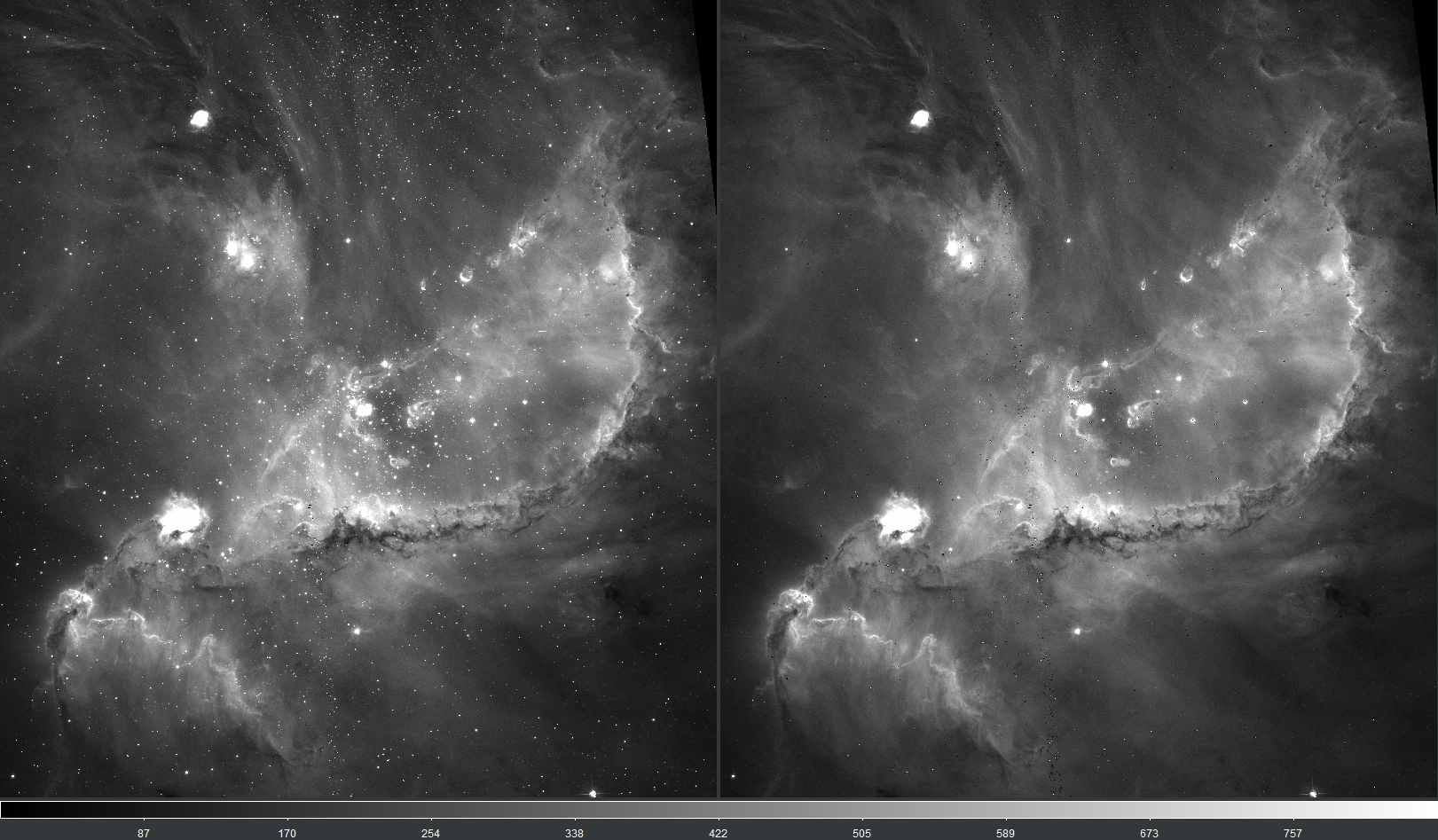}
\caption{Narrow-band H$\alpha$+[NII] HST ACS F658N images of NGC~346/N66. The images cover 3.3 $\times$ 3.3~arcmin$^2$ and are oriented with north up and east to the left. On the left is the original image and on the right is the image after removing the medium brightness stars. Images of saturated stars remain, but were removed for photometry. After numerically blanking the saturated stars, the data in the right image were used to measure H$\alpha$ fluxes and calibrate an archival image obtained with the ESAO VLT.}
\label{fig-acsn66}
\end{figure}

An archival HST ACS mosaic image in the F658N filter covers the main body of N66 as shown in \ref{fig-acsn66} (HST-GO-10248, P. I. A. Nota). We use the HST narrow band image of N66 to calibrate a wider field of view narrow-band H$\alpha$ image obtained from the ground. The contribution from stars to the narrow-band fluxes therefore needs to be removed from both images. 

The depth and superb angular resolution of HST allows stars to be clearly distinguished from nebular structure in the F658N image, which is  well resolved outside of the bright rim regions. We produced an approximation to a purely nebular image by cleaning stars from the ACS data. We preferred this approach for removing contributions from stars since the ACS observations of N66 do not include a filter that is free from significant flux contributions from the N66 nebula. 

The brightest stars are saturated in the ACS mosaic. We removed the saturated stars by setting all pixels in the stellar images above the maximum value of the diffuse nebular emission equal to the sky background. In this way, we blanked the bright stars. This process reduced the narrow band flux from the region of N66 covered by the HST image by 12\% after integrating over the field of view of the ACS data. 

We used PSF photometry for detecting and subtracting unsaturated stars brighter than m$_{F658N} \approx$ 20. The narrow-band image was imported into DAOphot to locate stars in the image that we detected assuming a stellar intensity profile FWHM of 2.5 pixels. The file with the positions of detected stars served as an input for DAOphot aperture stellar photometry using concentric apertures with pixel radii ranging from 2 to 12. Using data from the photometry we picked 20 stars with magnitude less than 20 to create an average model for the stellar point-spread function (PSF) required by the ALLSTAR sister program. ALLSTAR subtracts every stellar source in the image that matches the PSF model created in DAOphot. Using this process, we subtracted the majority of medium brightness stars in the narrow-band image. \ref{fig-acsn66} shows the ACS F658N mosaicked image before and after the PSF subtraction of stars.

After comparing the raw data and the star-subtracted data, we found that the intermediate brightness stars make up $\sim$5\% of the total narrow-band flux. The stellar contribution to the F658N image is therefore $\leq$ 17\%. We adopt the measured counts minus 17\% as a conservative estimate for the observed counts from the H$\alpha +$ [N II] emission lines. In correcting the H$\alpha$ flux, we adopt the \cite{Valerdi19} measurement of I([N II])/I(H$\alpha$)=0.02. Our total flux from the ACS image is therefore 81\% of the observed counts.

We converted corrected count rates to physical fluxes following HST calibration procedures. Using information in the header and data on the ACS system throughput with the F658N filter, we find an observed count rate of 1~DN/s = 1.69$\times$10$^{-16}$~erg~s$^{-1}$~cm$^{-2}$. The total observed flux from the region of N66 covered by the ACS therefore is approximately F(H$\alpha$)=4.1$\times$10$^{-10}$~erg~s$^{-1}$~cm$^{-2}$. This is a lower bound to the total observed F(H$\alpha$) from N66.

\begin{figure}[ht!]
\plotone{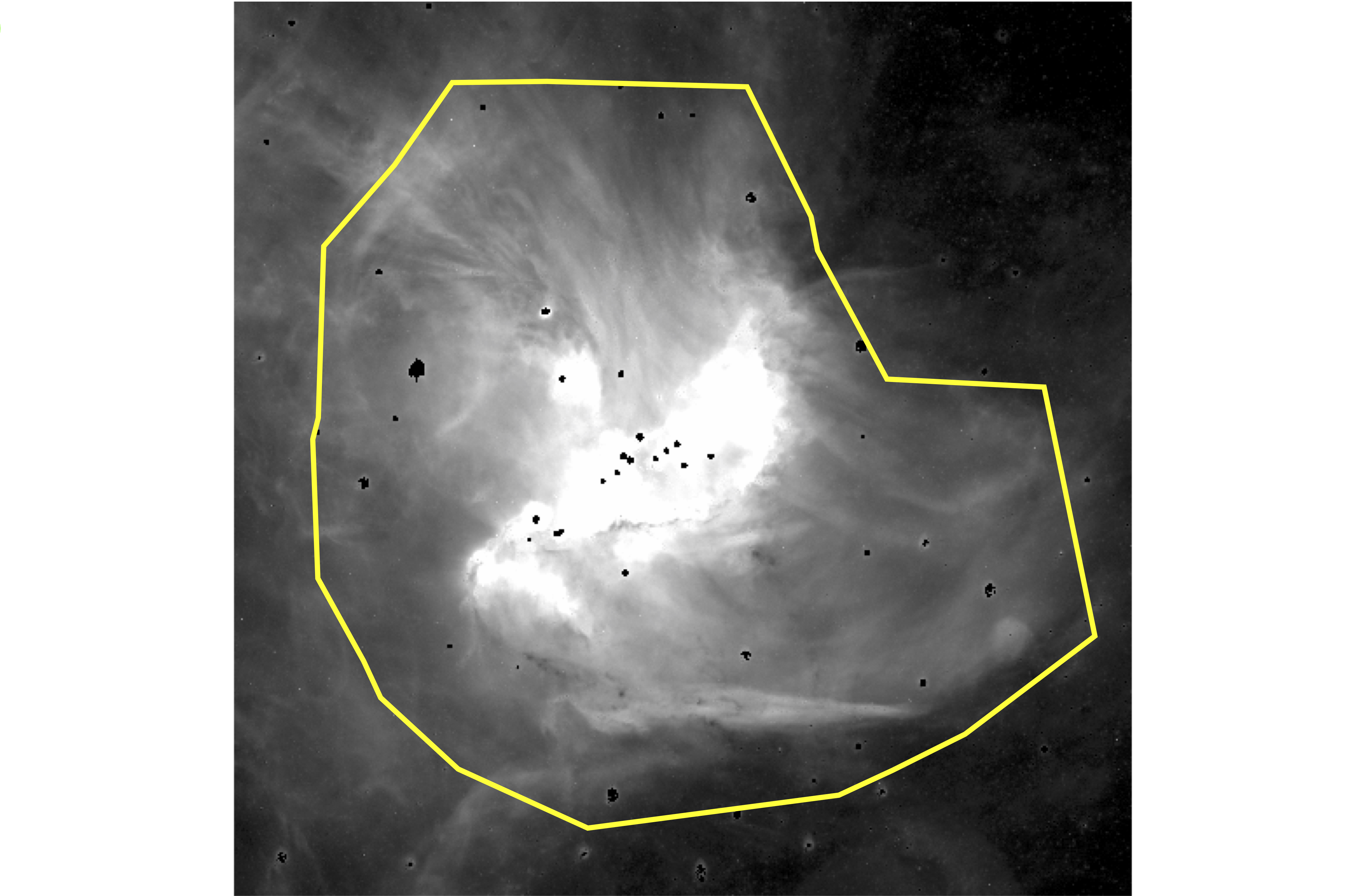}
\caption{Archival narrow-band H$\alpha$+[NII] VLT UT1 FORS~1 image of N66 obtained by E. Tolstoy. North is up and east is to the left. The image covers 6.8 $\times$ 6.8~arcmin$^2$. Intermediate brightness stars were subtracted and bright stars masked to produce this emission line image. The data in this image were used to measure H$\alpha$ fluxes within the region outlined by the lines.}
\label{fig-vltn66}
\end{figure}

More complete spatial coverage of N66 is provided by an archival narrow-band H$\alpha$ filter image of N66 from observations collected at the European Southern Observatory, proposal 63.N-0560, P. I. E. Tolstoy, using FORS~1 on VLT~UT1 (FORS.1999-08-20T05/10/28.741ff). FORS~1 has a field of view of 6.8 $\times$ 6.8~arcmin$^2$. Most of the stars were removed following the PSF fitting approach described above, leaving only a few stars that were too saturated for PSF photometry. We blanked the remaining stars by replacing all pixels above the 2500 counts per pixel, a level representative of the maximum counts per pixel in the nebula, with the intensity of the sky background. Figure~\ref{fig-vltn66} shows the FORS~1 image of N66 after cleaning the stars. 

The archival FORS~1 H$\alpha$ narrow-band image was calibrated by comparing sky-subtracted count rates in diffuse regions of N66 with data in the fluxed HST ACS image. This process yielded 1 FORS1 data number DN =1.88$\times$10$^{-19}$~erg~s$^{-1}$~cm$^{-2}$ with an uncertainty from the match between the two instruments of $\approx \pm$5\%. \cite{Caplan96} used a Fabry-Perot interferometer to isolate the H$\alpha$ line and, when we measure the flux in their 4.89~arcmin diameter aperture, our derived H$\alpha$ flux from the star-subtracted FORS~1 image agrees to within 2\% of their photoelectric measurements.

The total observed flux for N66 depends on the choice of boundaries for the N66 nebula. Our adopted region shown in Figure~\ref{fig-vltn66} gives the observed F(H$\alpha$)=6.8 $\times$10$^{-10}$~erg~s$^{-1}$~cm$^{-2}$ for the optically visible components of N66 following a correction of 5\% for [NII] emission. 

\subsection{Interstellar Obscuration}

\cite{Caplan96} and \cite{Valerdi19} discuss the reddening towards N66. Their data give an average observed intensity ratio of I(H$\alpha$)/I(H$\beta$)=3.11 $\pm$0.03. Using the nebular parameters adopted by \cite{Valerdi19}  with an electron temperature T$_e$=12500~K and electron density n$_e$=100~cm$^{-3}$, the reddening between H$\alpha$ and H$\beta$ is 0.12~mag. 

For a standard SMC extinction curve, the f(H$\alpha$) is reduced by 0.3 mag, consistent with the value found by \cite{Caplan96}. Therefore, the corrected H$\alpha$ flux based on the 0.3~mag of obscuration and a standard SMC extinction curve gives  f(H$\alpha$)$_0$=8.9 $\times$ 10$^{-10}$~erg~s$^{-1}$~cm$^{-2}$ and  L(H$\alpha$)=4.1 $\times$ 10$^{38}$~erg~s$^{-1}$ for N66.

Combining a mean H-atom density for the N66 HII region of n=100~cm$^{-3}$ \citep{Valerdi19} with the measured H$\alpha$ intensities yields an ionized gas volume filling factor of $\epsilon$(HII) = 0.1 along columns along the inner nebula containing the O-stars. The resulting gas column across the nebula is N$_H \approx$2$\times$ 10$^{21}$~cm$^{-2}$. This gives a minimum ionized gas mass of $\sim$10$^4$~M$_{\odot}$ for the inner $\sim$40~pc of N66. \cite{Hony15} find a total gas mass of 3$\times$10$^6$~M$_{\odot}$ and even after allowing for corrections for a significant mass fraction of ionized gas outside of the central half of the nebula, the ionized gas is only a small mass fraction of the N66 ISM complex. This suggests that ionization is most prevalent in low density zones within the interstellar cloud complex associated with N66. Many small ionization fronts are likely to exist within the N66 cloud complex. 

Using the above approximate numbers,  we expect A$_V$(N66) to be low, $\approx$0.05 through half of N66, and N66 will remain optically thin into the far ultraviolet \citep{Gordon03} as discussed by \cite{Hony15}.

\vspace{10mm}
\subsection{Ionization Rates}

Adopting the "on-the-spot" approximation where H-recombinations equal H-ionizations throughout the nebula and the nebular parameters from \cite{Valerdi19}, the ionizing photon flux Q$_0$ is approximately 2.2 times the photon luminosity in the H$\alpha$ line. Our data give L$_{\gamma}$(H$\alpha$)=1.4 $\times$10$^{50}$~s$^{-1}$. The ionization rate derived from our measurement of LH$\alpha$) in N66 is Q$_{0}$(Obs) $\approx$3 $\times$10$^{50}$~s$^{-1}$ for a distance of 62~kpc. We estimate the uncertainty from the measurements to be $\pm$10\% with another systematic uncertainty at the same level arising from the correction for absorption by dust for a total estimated uncertainty of Q$_{0}$(Obs) of $\pm$20\%.

\citet{Dufton19} provides a list of O-stars in N66 with spectroscopic measurements of temperatures and luminosities based on spectra obtained from terrestrial observatories. Using the calibration of ionization rates and spectral types for the SMC stars from \cite{Ramachandran19}, the Dufton sample gives an ionization rate of Q$_{0,*}$=2.4$\times$10$^{50}$~s$^{-1}$. Far ultraviolet spectra of stars in the crowded core of the main NGC~346 cluster were obtained by program HST-GO-15112 (P. I. L. Oskinova), These results are analyzed in Rickard {\it etal} (in preparation) and provide an additional stellar ionizing flux of $\approx$2$\times$10$^{49}$~s$^{-1}$ to give an estimated total Q$_{0,*} \gtrsim $2.6$\times$10$^{50}$~s$^{-1}$. This is a lower limit to the stellar ionization rate as undetected massive binary stellar companions are likely to be present but are not fully accounted for.

\subsection{SALT Spectra}

A long slit spectrum of N66 was obtained on August 3, 2021 with the Robert Stobie Spectrograph (RSS \cite{Burgh03}) on the Southern Africa Large Telescope (SALT,\cite{Buckley06}) . The spectral resolution was $\approx$900 with the spectra covering from [OII] $\lambda$3727\AA\ to the H$\alpha +$ [NII] emission lines. The RSS observation used a slit width of 1.0~arcsec and length of 8~arcmin. 

Here we present an initial analysis of line ratios obtained from flat-fielded and bias-corrected data. The spectra for this study were calibrated against the relative flux line intensities measured by \cite{Valerdi19}. This approach is effective since emission line ratios show little spatial variation across the main body of N66. In particular, our data show that the observed ratio of I(H$\beta$)/I(H$\alpha$) remains constant to better than 10\% along both of the SALT slits. This is an indication that the level of interstellar obscuration is quite constant across much of N66 and that the structure of the optical nebula is not strongly affected by dust obscuration. We show the slit orientation and discuss results obtained from the spectrum in \S3.2.

\subsection{Mid-Infrared Luminosity} \label{subsec:n11_midir}

\begin{figure}[ht!]
\plotone{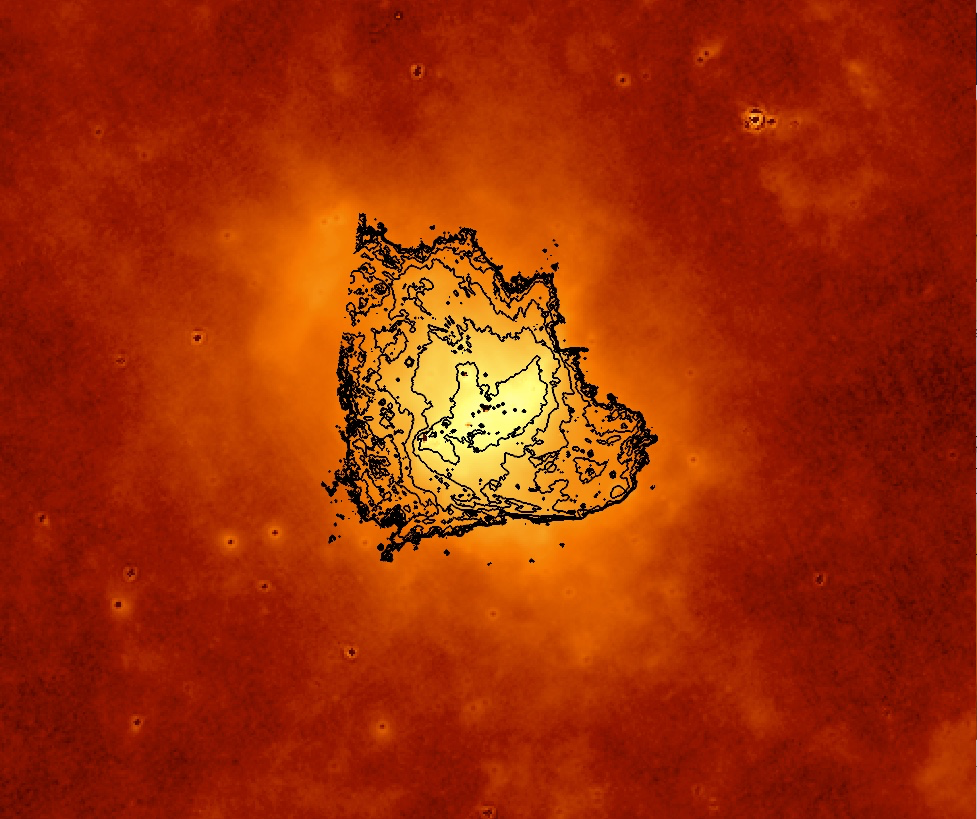}
\caption{N66 region from 24~$\mu m$ image obtained by SAGE-SMC covering 19.7 $\times$ 17~arcmin$^2$ oriented with north at the top and east on the left. Contours are from the FORS~1 image of N66 and are limited in extent by the FORS~1 field of view. Note the similarity between the spatial patterns of H$\alpha$ and 24~$\mu m$ emission and the structural continuity to the outer regions of the N66 interstellar complex as seen from the wider field-of-view {\it Spitzer Space Telescope} data.}
\label{fig-n66sage24}
\end{figure}

Aside from filamentary ionization boundaries associated with dense gas that cover only a small fraction of the surface of N66 (see \S3.2), the nebula is optically thin in the mid-infrared \citep{Hony15}. Mid-infrared emission therefore comes from the entire interstellar matter complex associated with N66. Observations of the mid-infrared structure of N66 therefore provide a more complete map of the nebula than the possibly ionization-bounded H$\alpha$ images. Using data from SAGE-SMC taken with the SST \cite{Gordon11}, we made a 24~$\mu m$ cutout of the N66 region. A comparison between the VLT H$\alpha$ star-subtracted and 24~$\mu m$ SST data is shown in Figure \ref{fig-n66sage24}. The 24~$\mu m$ intensity generally follows that of the main HII region at a flux level greater than approximately 2.3~MJy~sr$^{-1}$. At an intensity of 0.5~MJy~sr$^{-1}$, 2.5 times the $\approx$0.2~MJy~sr$^{-1}$ background, we see in Figure \ref{fig-n66sage24} that the H$\alpha$ and 24~$\mu $ emission from warm dust are similar. This is a natural result from the transparency of the interstellar matter in N66 that allows for extensive UV heating of dust and ionization extending through much of this region. 

We photometered N66 on the 24~$\mu m$ SMC-SAGE image within an isophote following the 0.6~MJy~sr$^{-1}$ intensity level using a background of 0.2~~MJy~sr$^{-1}$. The mean intensity in the aperture, after correcting for minor emission contributions from compact sources, is 4.5~MJy~sr$^{-1}$ in an area of 2.77 $\times$10$^5$~arcsec$^2$ (6.52 $\times$10$^{-6}$~ sr). The mean isophotal 24~$\mu m$ diameter of N66 based on its area is $\sim$150~pc. 

We assign an uncertainty of $\pm$10\% to the integrated F(24~$\mu m$)=29~Jy for N66 set by the basic calibration and the choice of aperture. This flux corresponds a total 24~$\mu m$ luminosity of $\nu$L(24~$\mu m$) =1.7$\times$10$^{39}$~erg~s$^{-1}$. Only $\approx$ 10\% of the $\nu$L(24~$\mu m$) from N66 is produced in the high brightness core around the main NGC~346 star cluster. This small mid-IR luminosity fraction indicates that only a small area within N66 is sufficiently dense to be optically thick and efficiently absorb FUV stellar radiation.  Most of the 24~$\mu m$ luminosity originates from diffuse emission within the ionized nebula. 

Most of the radiation produced by young stars in NGC~346 is radiated at wavelengths above the Lyman edge. Low angular resolution far ultraviolet (FUV) photometry by \cite{Nandy79} and \cite{Cornett97} give L(FUV)$\sim$10$^7$~L$_{\odot}$ or about 10 times L(24)~${\mu}$. This estimate is uncertain at the factor of order a few due to the unknown distribution of interstellar extinctions and low angular resolution of the FUV data. Despite these uncertainties it is clear that the majority of the FUV stellar luminosity produced by massive stars in NGC~346 streams through the low optical depth N66 nebula and into the surrounding interstellar medium. 

\vspace{5mm}
 \section{Discussion}

\subsection{Integrated Star Formation Rates}\label{subsec:sfr}

The SFR can be estimated by assuming that star formation in N66 occurs over a time span that is short compared to the lifetimes of the massive stars. In this case, we can use an instantaneous stellar population model to derive the expected stellar mass, and estimate the SFR by dividing the predicted stellar mass by the empirically derived ages of the massive stars. We computed an instantaneous star formation model using the current public version of the Starburst99 V7.01 code \footnote{https://www.stsci.edu/science/starburst99/docs/default.htm} \cite[see][]{Leitherer99}. We adopted a standard form of the Kroupa stellar initial mass function (IMF) implemented in Starburst99 with a low mass IMF slope break at 0.5~M$_{\odot}$, and used the Geneva stellar model option with Z=0.002 with an upper mass limit of 60~M$_{\odot}$. Our model was run for a total stellar mass of 10$^6$~M$_{\odot}$ to minimize stochastic effects. We scaled the model results to the N66/NGC~346 system from the ratios of the observed to predicted ionizing photons rate. The models give log(Q$_0$(model))=52.45$\pm$0.05 for ages of 1-4~Myr following the instant formation of the idealized stellar system with a mass of 10$^6$~M$_{\odot}$. This is a factor of 94 larger than our observed ionizing flux of log(Q$_0$(N66))=50.5, and we  therefore infer a stellar mass of $\approx$10$^4$~M$_\odot$ for the NGC 346 ionizing cluster.  Estimates for the mass of NGC~346 from observations of resolved stars with mass of $>$0.8~M$_{\odot}$  give a lower limit of 1.2$\times$10$^4$~M$_{\odot}$ for the main stellar concentration in an 88$\times$88~pc$^2$ region and a total stellar mass that is likely to be about twice this value \citep{Sabbi07}. This difference in mass derived from the massive and low mass stars may be associated with the lack of a spatial correlation between high and low mass stars in NGC~346 that could result from different evolutionary processes in the various subregions of NGC~346 \citep{Sabbi11}.

The SFR based on the massive star population age of 
t$_{SF} \approx$3~Myr \cite[e.g.,][]{Sabbi07, Cignoni10}\footnote{The overall time scale for star formation in the NGC~346 system, however, may be longer than 3~Myr, with lower mass stars possibly having formed somewhat earlier \citep{Sabbi11, DeMarchi11, Hony15} which would reduce the estimated SFR} is approximately 0.003(3$\times$10$^6$ yr/t$_{SF}$)~M$_{\odot}$~yr$^{-1}$, similar to that derived by \cite{Hony15}. When averaged over the full extent of the interstellar cloud complex as defined by the N66 nebula, $\Sigma_{SFR} \approx$ 1 $\times$10$^{-7}$~M$_{\odot}$~pc$^{-2}$~yr$^{-1}$, which is in agreement to within a factor of a few of the $\Sigma_{SFR}$ found by \cite{Cignoni11}. 

The N66/NGC~346 system is well-known for presenting an opportunity to explore connections between the high and low ends of the stellar mass distributions in a low metallicity environment. In the above calculation, the total stellar mass is inferred by adopting a standard form for the stellar IMF and compared this with stellar mass estimates derived from star counts using HST images. We expect that the properties of the low mass stellar populations in NGC 346, including their ages, will be further refined from deep imaging with the James Webb Space Telescope. The relationships between low and high mass star formation can then be more directly determined as functions of time and space to gain a better understanding of the development of the stellar mass distribution in this region.

The situation in N66 is typical of individual HII regions where Q$_0$ is extremely sensitive to the presence of the most massive stars and their evolutionary status. \citep[e.g.,][]{Ramachandran19}. Even under the unphysical assumption that binary stars are not a factor in producing Q$_0$, the prediction of L(H$\alpha$) under the assumption of optically thick, ionization bounded nebula will be uncertain by a factor of a few for HII regions in the luminosity range of N66 \citep[e.g.,][]{Fumagalli11, Orozoco22}. The situation is further complicated when binary evolution is included \citep{Eldridge12}. 

Recent observations of the massive stars in the R136 star cluster at the heart of the LMC's giant Tarantula nebula HII region present another possible factor. R136 contains stars with M$>$100~M$_{\odot}$ that dominate Q$_0$ from the cluster \citep{Schneider18,Bestenlehner20}. Some of these stars may be produced by mergers in dense environments \citep{Banerjee12}, which could add an environmental connection to the relationship between L(H$\alpha$) and stellar mass.

The results from N66 illustrate some of the problems in determining SFRs in low metallicity environments. Since main sequence O-stars have weak winds, mechanical feedback is delayed until massive stars leave the main sequence. This effect leads to quiescent HII regions, such as N66, where mechanical feedback from massive, pre-supernova stars may be reduced. This raises the possibility that low mass star formation processes could differ from those in systems where stellar winds produce turbulent, higher pressure HII regions. In such environments, binary evolution may be especially important in producing high temperature stripped stellar cores that can enhance ionization rates.

The reduced dust opacity decreases the local absorption of FUV luminosity. This effect lowers mid-IR luminosities and SFRs derived using standard relationships \citep[e.g.,][]{Kennicutt12}. The combination of these factors suggests that SFR measurements of low metallicity galaxies based on L(H$\alpha$) or $\nu$L$_{\nu}$(22~$\mu$m) need to be carefully calibrated against nearby low metallicity galaxies where young stellar masses can be directly observed to obtain a fuller understanding of possible uncertainties.

The results of the measurements presented in this paper, with associated uncertainties, are summarized in Table \ref{tab:n11params}.

\begin{deluxetable*}{lll}
\tablenum{1}
\tablecaption{Properties of the SMC-N66 H~II Region \label{tab:n11params}}
\tablewidth{0pt}
\tablehead{
\colhead{Parameter} & \colhead{Measurement} & \colhead{Uncertainty}}
%\decimalcolnumbers
\startdata
Distance & 62~kpc\\
Diameter & $\sim$150~pc & see text\\
H$\alpha$ flux & f(H$\alpha$)$_0$ = 8.9 $\times$ 10$^{-10}$erg~s$^{-1}$ & $\pm$10\%\\
H$\alpha$ luminosity & L(H$\alpha$) = 4.1 $\times$ 10$^{38}$~erg~s$^{-1}$ & $\pm$10\%\\
HII mass & $\sim$10$^4$~M$_{\odot}$ & see text\\
Mid-IR Luminosity & $\nu$L$_{\nu}$(24~m$\mu$) =1.7$\times 10^{39}$~erg~s$^{-1}$ & $\pm$10\%\\
Far-UV luminosity & L(FUV) $\approx$4$ \times 10^{40}$~erg~s$^{-1}$ & $\pm \approx$50\% \\
HII ionization rate & Q$_0$(obs) $\approx$3$ \times 10^{50}$~s$^{-1}$ & $\pm$20\%\\
Stellar ionization rate & Q$_{0,*} \gtrsim$2.6$\times$10$^{50}$~s$^{-1}$ & see text\\
Estimated star formation rate & SFR $\sim$3 $\times$ 10$^{-3}$~M$_{\odot}$~ yr$^{-1}$ & see text
\enddata
\tablecomments{The uncertainty in f(H$\alpha$) is 7\% from the photometric calibration and residual star images and an equal factor for uncertainty in extinction for an estimated 10\% uncertainty that then also applies to L(H$\alpha$). We do not propagate the effect of a potential $\sim$5\%  uncertainty in distance in our distance-dependent estimates.}
\end{deluxetable*}

\vspace{-5pt}
\subsection{Structure of N66}

The N66 nebula shows structural features extending over a variety of spatial scales \citep[e.g.,][]{Ye91,Hony15}. The extent of an ionization bounded HII region is determined by the photoionizing luminosity in combination with the structure of the associated ISM. In ionization bounded nebulae, the size of the nebula is determined by the photoionizing luminosity, and substantial amounts of neutral or molecular matter will exist beyond the ionization fronts. The compression of gas along the ionization fronts produced by the expanding ionized nebula leads to complex, relatively dense zones where ionization levels and emission line intensities vary rapidly with position across the front \citep[e.g.,][]{Hester96, Osterbrock06}. 

However, in the ideal case of matter-bounded nebulae, the size of a bounded HII region is set by the distribution of gas as ionization throughout the volume of over-dense ISM, and the stellar ionizing flux escapes into the surroundings. HII regions often combine both features, being ionization bounded in some directions and leaking photoionization in the directions where the gas density is low \citep[see, e.g.,][]{Pellegrini12}.

Well-defined ionization fronts in N66 are bright boundaries to the H$\alpha$ emission that form a loop-like structure around the main NGC~346 cluster extending in an approximately east-west direction. A second clear ionization boundary can be seen running in a north-south direction to the north of the main NGC 346 cluster, and a third, arc-like boundary defines the southern edge of N66. The region of the northern ionization front is suspected to be the site of a collision between molecular clouds  that is fostering star formation \citep{Neemlamkodan21}. 

\begin{figure}[ht!]
\plotone{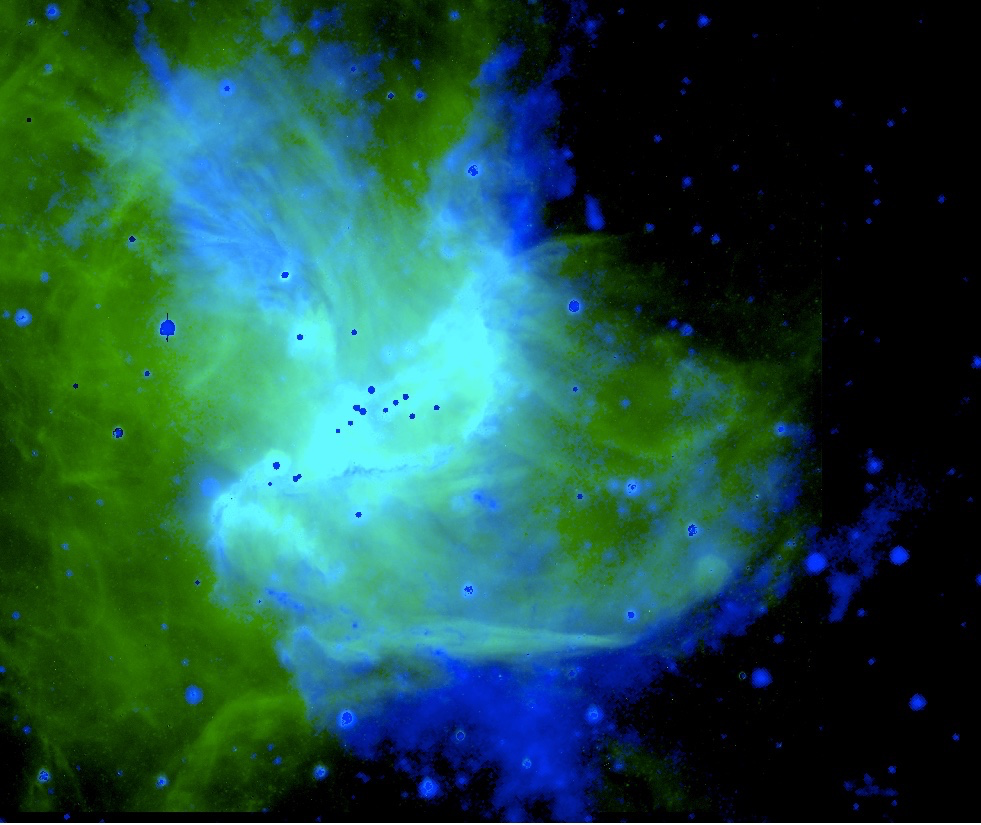}
\caption{This figure compares the 8~$\mu m$ emission from SMC-SAGE, primarily arising from PAH emission in dense boundaries between ionized and neutral gas (blue) and H$\alpha$ emission from the ionized N66 nebula (green). While the nebula appears to be constrained by dense interstellar material to the south and along the edge of the northern plume, other areas do not show the presence of well-defined nebular boundaries produced by dense interstellar matter. Scale and orientation are the same as in \ref{fig-vltn66}.}
\label{fig-n668mic}
\end{figure}

The ionization fronts also stand out in infrared images obtained by ISOCAM and SMC-SAGE \citep{Contursi00,Gordon11}, especially in the light of the 8~$\mu m$ PAH emission features \citep[e.g.,][]{Sabbi11, Hony15}. Figure~\ref{fig-n668mic} shows the 8~$\mu m$ emission associated with N66. Dense ISM that protects PAH emitters from stellar UV is required to produce strong 8~$\mu m$ PAH emission in low metallicity systems such as the SMC \citep{Oey17}. Thus, these regions only occur in N66 at the surfaces of dense interstellar matter, i.e., along the locations of well defined ionization fronts where high column densities provide protection from intense stellar UV radiation.

The comparison of 8~$\mu m$ and H$\alpha$ emission maps in Figure~\ref{fig-n668mic} indicates that N66 is strongly ionization bounded by dense ISM at its southern edge and over a substantial part of the lower half of the nebula. A pair of strong ionization fronts also define the boundaries of the northern extension of N66, a region where the nebula has a filamentary structure. The organized nature of the 8~$\mu m$ emission suggests that the dense ISM is organized on relatively large scales around N66, as discussed by \cite{Rubio00,Contursi00,Hony15} who present CO and H$_2$ emission maps of N66. Dense ionization boundaries are not well defined in some directions, e.g. to the north, where ionizing radiation may escape and where molecular gas has not been detected. However, we emphasize that due to the low dust optical depths in the SMC, an absence of 8~$\mu m$ emission does not necessarily imply a lack of ionization fronts \citep{Oey17}.

Within the HII region, the emission on moderate scales of ~10-20~pc is often filamentary, especially in the northern extension of the nebula where filaments generally have a north-south orientation. Filament widths are a few parcsecs, and intensity variations run at the 10\% level. To the south of the NGC 346 cluster, extended filaments are less frequent and include curved structures. \cite{Ye91} interpret the filamentary structures as arising from the matter on the edges of a ``champagne flow", i.e. a supersonic outflow from the core of the HII region. A champagne flow occurs when the higher pressure ionized gas escapes from a confining cooler cloud, usually at supersonic velocities and over a limited range of solid angles \citep[e.g.,][]{Tagle79}.  However, \cite{Smith12} finds that the north and other regions of N66 are kinematically quiescent; the emission line widths in Smith's position 3 are comparable to the sound speed in the ionized gas. The currently available data do not provide evidence for a dynamical origin for the filaments. 

Since Balmer line emissivity in an HII regions scales with the electron density n$_e$ and path length L as n$_e^2$L, compact, bright regions, such as the filaments or ionization fronts, either have a long L or higher n$_e$ as compared with their surroundings. Since the filaments have small widths, they are unlikely to be distinguished by large L and probably mark locations of denser ionized gas in N66. The alternative, that the filaments are produced by foreground dust extinction, can be excluded based on the constant ratios of Balmer emission lines across filaments seen in our (see below) and and \cite{Relano02} spectra.   

In \S~2.2 we found that the ionized gas mass comprises $\sim$1\% of the total mass in the N66 cloud complex. Evidently, substantial amounts of unionized interstellar matter exists around and within N66. The filamentary structure could therefore arise from ionization fronts within the nebula along the surfaces of embedded neutral or molecular gas. In this case, the filamentary nature of these regions reflects the intrinsic projected density structure of the N66 cloud.

A further perspective on the relationship between the ISM over-density associated with N66 and the HII region comes from a comparison between the H$\alpha$ and  60~$\mu m$ images. The 60~$\mu m$ emission comes from dust warmed by the UV radiation that, due to the low dust content, is not trapped within the N66 cloud. The 60~$\mu m$ image therefore shows the full extent of the interstellar cloud complex associated with N66. If N66 is fully ionization bounded, then we expect the cloud complex to extend beyond the HII region. The converse would be true in a matter-bounded case where ionization could extend to the limits of the cloud complex.

\begin{figure}[ht!]
\plotone{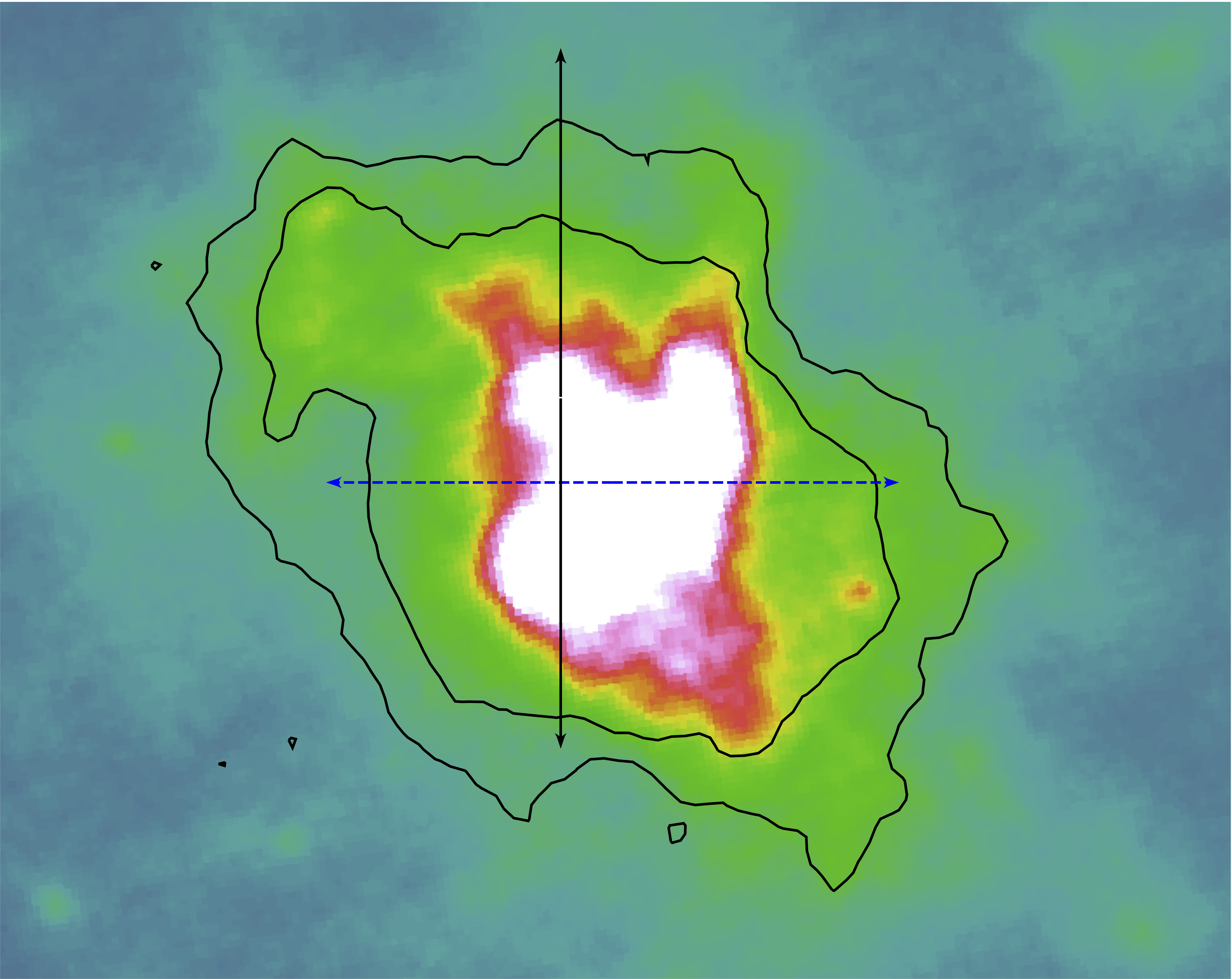}
\caption{This figure shows the N66 region in dust emission at 70~$\mu m$ from SMC-SAGE. North is at the top and east is to the left. The contours are from the 24~$\mu m$ image at levels of 1 and 2 MJy~sr$^{-1}$, which illustrates the similar structure of the N66 region at mid and far infrared wavelengths. The north-south solid line marks the 8-arcmin length of the north-south slit used to obtain emission line ratios from our SALT spectra. Line ratios change in the region north of the outermost contour. The dashed lines show the location of the 6.8~arcmin length east-west slit A from \citet{Valerdi19}.}
\label{fig:n66_70_24mic}
\end{figure}

A comparison between Figure \ref{fig:n66_70_24mic} and Figure \ref{fig-n66sage24} demonstrates that the structure of the N66 cloud, as defined by infrared emission, is similar at wavelengths of 24~$\mu$m and 60~$\mu$m. This pattern is repeated at longer wavelengths where the structure of the N66 complex is similar even when mapped in the 1.1~mm dust continuum by \cite{Takekoshi17} \citep[see also][]{Hony15}. At long wavelengths, the size of the N66 complex is not substantially influenced by changes in dust temperature. The consistency of the brightness structure of N66 from the mid-IR to the millimeter indicates that the mid-infrared structure is mainly driven by the distribution of matter rather than the properties of the radiation field responsible for heating the dust. The 60~$\mu$m image therefore provides an approximate map of the density of matter in the interstellar cloud complex associated with N66.

 Figure~\ref{fig-n66sage24} shows the shape of the main HII region parallels that of the cloud complex and extends close to the edges of the density distribution on the north and eastern sides. The coincidence between the extent of the host cloud complex and ionized gas in these directions suggests that ionizing radiation reaches to the limits of the denser ISM in some directions. N66 may be matter-bounded in some directions, which would allow ionizing radiation to escape into the surroundings.

Long slit optical spectra extending beyond the main body of N66 were obtained in an east-west direction by \cite{Valerdi19} and by ourselves to the north using SALT (see slit orientation shown in Figure \ref{fig:n66_70_24mic}). The Valerdi {\it et al.} data and our data demonstrate that ionized gas extends over the full extent of the N66 cloud complex. Both sets of spectra show a decrease in the flux ratios of F([OIII])/F([OII]) and F(H$\beta$)/F([OIII]) only at the very ends of their east-west slits. This trend is especially clear approximately 200~arcsec west of the center of the main NGC~346 cluster. 

These changes in line ratios are as expected when direct photoionization by the stellar radiation field is dropping at an ionization front \citep[e.g.,][]{Hester96}. However, there is no clear increase in the strength of [SII] relative to the forbidden oxygen lines in the Valerdi {it et al.} spectra as would be expected if the slits had crossed the ionization front. The presence of enhanced [SII] emission around the edge of the nebula is discussed by \cite{Pellegrini12} as evidence that the N66 traps all of its ionizing radiation. However, \cite{Reid06} also present an image of the N66 region showing the intensity ratio of [SII]/H$\alpha$  suggesting that the regions with strong [SII] emission marking locations of well-defined ionization fronts do not completely encircle the nebula. This opens the possibility that for the escape of ionizing radiation in some directions.

\begin{figure}[ht!]
\plotone{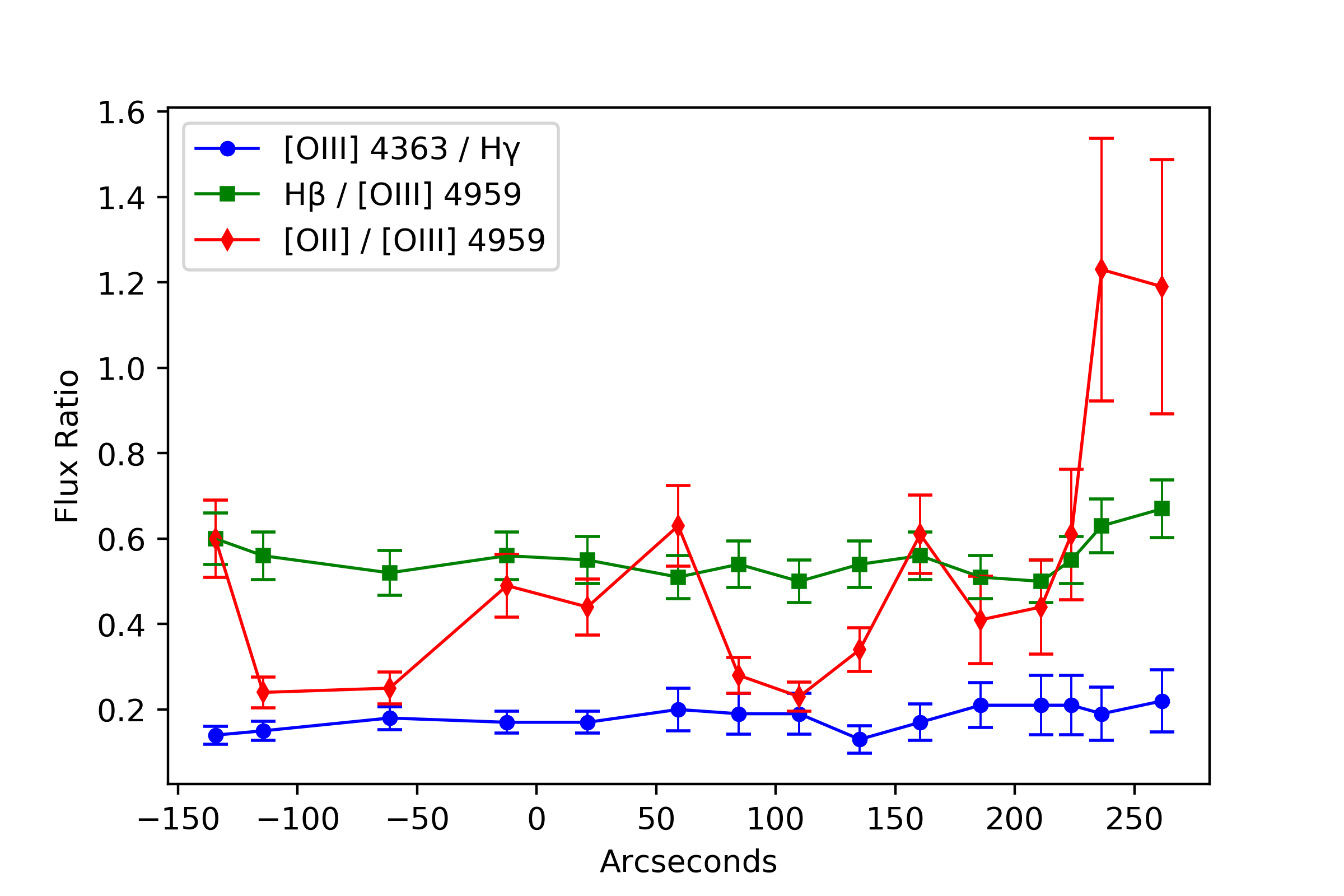}
\caption{Emission line ratios derived from our RSS SALT spectra are shown running from south to north. The transition away from direct ionization begins at the extreme northern end of our slit which is also the region where the rate of decline in the H-Balmer line intensities flattens.}
\label{fig-rss}
\end{figure}

The results for line ratios derived from the analysis of our SALT spectra are shown in Figure~\ref{fig-rss}. A pattern similar to that by \cite{Valerdi19} is present with a degree of ionization based on the ratio of I([OII])/I([OIII]) sharply increasing beyond a north distance of 200~arcseconds. As shown in Figure~\ref{fig:n66_70_24mic}, this region is at the northern edge of the main gas structure associated with N66, so the nebula may be matter-bounded in this region. The possibility therefore remains that we are seeing a diffuse ionization front that bounds the N66 HII region as suggested by \cite{Pellegrini12}.

Alternatively, as the intensity of the nebular emission declines, we may be observing a larger contribution from diffuse ionized gas in the SMC. Emission line ratios in the low density outer regions of the N66 gas complex could therefore be a superposition of decreasing contributions from material in N66 combined with diffuse ionized background emission from the SMC. Diffuse, ionized gas is expected to have a lower ionization level than the material in N66 \citep[e.g., ][]{Hunter90}, and would therefore produce an increase in the observed I([OII])/I([OIII]). If this situation occurs, then the intensity of the H-Balmer emission lines as a function of position along the slit should become flatter as the diffuse emission becomes dominant and the electron temperature would not need to drop as expected in a classical ionization front.

We observe a flat distribution of H-Balmer emission and temperature-sensitive [OIII] intensity ratios over approximately the northern 50~arcseconds of the SALT slit, consistent with a transition to emission from the diffuse ionized background. In this case, ionizing photons would be locally escaping from N66. However, given the limited coverage provided by the existing data, further observations are required to clarify whether ionization escapes through channels in the boundaries of the N66 nebula. Spectroscopic studies of the boundary regions of the N66 cloud complex, as determined from mid-infrared dust emission images, have the potential to distinguish between bounding by diffuse ionization fronts versus ionization escape combining with a contribution from the SMC's diffuse ionized background to produce the observed line intensities. In the latter case, kinematic measurements could be useful in distinguishing line emission associated with N66 from that produced over much larger volumes of the SMC.

\section{Conclusions}

Photometry of narrow-band nebular H$\alpha$ images of N66 give an observed total flux of F(H$\alpha$) = 6.8 $\times$10$^{-10}$~erg~s$^{-1}$~cm$^{-2}$. Our results, calibrated from narrow-band HST imaging, are in good agreement with the precision aperture photometry by \cite{Caplan96}. When corrected for interstellar obscuration by a  factor of 0.3 based on the Balmer decrement, the H$\alpha$ luminosity of N66 is L(H$\alpha$)=4.1 $\times$ 10$^{38}$(D$_{SMC}$/62~kpc)$^2$~erg~s$^{-1}$.

The estimate of stellar ionizing photons has recently been updated through HST UV spectroscopy of stars in the crowded core of the main NGC~346 cluster. The resulting stellar H-ionization rate of Q$_{0,*}  \approx $2.6$\times$10$^{50}$~s$^{-1}$ is consistent with the nebular ionization rate of Q$_0$(obs) $\approx$3 $\times$10$^{50}$~s$^{-1}$. Even though uncertainties remain (e.g., understanding the role of dust absorption, role of binary stars), N66 is approximately in ionization equilibrium with the massive stars in the NGC~346 star clusters, consistent with \cite{Pellegrini12} result that the N66 interstellar complex, while optically thin to dust absorption, is largely optically thick to ionizing stellar photons. 

The young stellar mass that we derived from the N66 ionization rate is a factor of 2-3 times lower than the observed mass of young stars in the region. This difference is somewhat surprising since N66 may be near the peak of its production of ionizing photons, and is still producing lower mass stars. Therefore, the model stellar mass derived from an assumed instantaneous formation of a full stellar mass spectrum would be expected to under-predict the total stellar mass. 

This issue is possibly connected to the differences in the spatial distributions of high and low mass stars in NGC~346 that has been noted by several authors. Perhaps we are observing a recent massive star forming event superposed in space on more extensive regions dominated by the formation of low mass stars? Forthcoming further explorations of the development of stellar populations in this region are needed to understand the patterns of star formation in space and time in NGC~346. Statistical sampling of the most massive stars is also important in deriving Q$_{0,*}$ and associated stellar masses. This can be seen in that the O2~V star NGC 346 MPG355 (see \cite{Dufton19}) alone provides about 1/3 of the observed stellar ionizing photons in N66.

The combination of H$\alpha$ and infrared images from SMC-SAGE indicate that N66 contains regions of denser interstellar matter that are arrayed in east-west and northern directions within the larger interstellar complex associated with N66. The dense regions are embedded in an ISM structure with a diameter of $\approx$150~pc. Most of this region is transparent to the FUV luminosity of the massive stars above the Lyman edge, so far-ultraviolet radiation above the Lyman edge from the massive stars in NGC 346 largely escapes from N66 and illuminates its surroundings. 

Since the N66 ISM complex is mostly optically thin to the UV radiation above the Lyman continuum responsible for dust heating, the mid-and far-infrared structures are similar and provide a map of the gas distribution. Ionized gas is nearly coincident with the extent of the N66 region as given by the SMC-SAGE mid- and far-infrared images, but comprises only a small fraction of the mass of the N66 interstellar complex. Optical spectra that sample the outer parts of N66 indicate that ionization permeates almost the entire ISM complex. Ionization most likely occurs in lower density regions throughout the cloud complex. The distinctive filamentary structures seen in the outer parts of N66 could arise from local ionization fronts on multiple boundaries between ionized and neutral gas throughout the N66 interstellar cloud complex. In this situation, the filaments would be due to the intrinsically filamentary projected density distribution within the N66 cloud.

Even though N66 is in near ionization balance with its massive stars, some ionization leakage may occur from lower density zones. Clearly defined, large-scale ionization boundaries associated with high density ISM are only seen along a small portion of N66's periphery. Optical spectra suggest that diffuse Lyman continuum could be leaking in several directions where the N66 HII region appears to be matter-bounded, although ionization channels where stellar ionizing photons directly escape are not detected. 

Several of the distinctive properties of N66, including weak main sequence O-star stellar winds and low interstellar dust opacity, are intrinsic to low metallicity HII regions. These features, as well as intrinsic properties of its massive stars, especially the nature and frequency of binaries, affect the SFR indicators based on the mid-infrared and H$\alpha$ luminosities. The derived SFRs from these measurements can be more uncertain than in better studied solar metallicity HII regions. Additional studies of the structure of N66 are worthwhile as a way to better understand how giant star forming regions influence their environments in low metallicity galaxies.

 \begin{acknowledgements}
 We thank the referee for constructive comments that improved this paper. 
 This research is based on observations made with the NASA/ESA Hubble Space Telescope obtained from the Space Telescope Science Institute, which is operated by the Association of Universities for Research in Astronomy, Inc., under NASA contract NAS~5–26555. These observations are associated with programs GO-10248 and GO-15112. This work uses observations obtained with the Southern African Large Telescope (SALT),
programme 2021-1-SCI-023. We thank the SAGE team for providing the community with excellent archival data sets. Partial support for this research was supplied by by the National Science Foundation's REU program in Astrophysics through NSF award AST-1852136.  A.Y.K. acknowledges support from the National Research Foundation (NRF) of South Africa. RK gratefully acknowledges partial funding support from the National Aeronautics and Space Administration under project 80NSSC18K1498, and from the National Science Foundation under grants No 1852136 and 2150222. JSG and RK express their appreciation to the University of Wisconsin-Madison for partial support of this research and access to SALT. VR acknowledges support by the Deutsche Forschungsgemeinschaft (DFG-German Research Foundation) in the form of an Emmy Noether Research Group (grant number SA4064/1-1, PI Sander).
 \end{acknowledgements}
 
\vspace{5mm}
\facilities{HST(ACS,STIS) STSCI(MAST) ESO(Science Archive), SALT(RSS), SST(IRAC, MIPS)}
\software{IRAF PYSALT, python3 ds9 STARBURST99 V7.0.1}

\bibliography{n66_refs}{}
\bibliographystyle{aasjournal}

\end{document}